\documentclass[superscriptaddress,prl,twocolumn,nofootinbib,showpacs,preprintnumbers,floatfix]{revtex4}

\usepackage{amsfonts}
\usepackage{mathrsfs}
\usepackage{amssymb}
\usepackage{amsmath}
\usepackage{graphicx,epsfig}

\def\bwt{\begin{widetext}}

\def\ewt{\end{widetext}}

\def\be{\begin{equation}}

\def\ee{\end{equation}}

\def\bea{\begin{eqnarray}}

\def\eea{\end{eqnarray}}

\def\bean{\begin{eqnarray*}}

\def\eean{\end{eqnarray*}}

\def\bary{\begin{array}}

\def\eary{\end{array}}

\def\bit{\begin{itemize}}

\def\eit{\end{itemize}}

\def\su5u1{SU(5) \times U(1)}

\def\fsu5u1{SU(5) \times U(1)'}

\def\so10{SO(10)}

\def\sq20{SO(10) \times SO(10)}

\usepackage{amssymb}

\begin{document}

\setlength{\parskip}{0cm}

\title{ The Golden Point of No-Scale and No-Parameter ${\cal F}$-$SU(5)$ }

\author{Tianjun Li}

\affiliation{George P. and Cynthia W. Mitchell Institute for
Fundamental Physics, Texas A$\&$M University, College Station, TX
77843, USA }

\affiliation{Key Laboratory of Frontiers in Theoretical Physics,
      Institute of Theoretical Physics, Chinese Academy of Sciences,
Beijing 100190, P. R. China }

\author{James A. Maxin}

\affiliation{George P. and Cynthia W. Mitchell Institute for
Fundamental Physics, Texas A$\&$M University, College Station, TX
77843, USA }

\author{Dimitri V. Nanopoulos}

\affiliation{George P. and Cynthia W. Mitchell Institute for
Fundamental Physics,
 Texas A$\&$M University, College Station, TX 77843, USA }

\affiliation{Astroparticle Physics Group,
Houston Advanced Research Center (HARC),
Mitchell Campus, Woodlands, TX 77381, USA}

\affiliation{Academy of Athens, Division of Natural Sciences,
 28 Panepistimiou Avenue, Athens 10679, Greece }

\author{Joel W. Walker}

\affiliation{Department of Physics, Sam Houston State University,
Huntsville, TX 77341, USA }



\begin{abstract}

The ${\cal F}$-lipped $SU(5)\times U(1)_X$ Grand Unified Theory (GUT)
supplemented by TeV-scale vector-like particles from ${\cal F}$-theory,
together dubbed ${\cal F}$-$SU(5)$, offers a natural multi-phase unification process
which suggests an elegant implementation of the No-Scale Supergravity boundary conditions
at the unification scale $M_{\cal F} \simeq 7 \times 10^{17}$~GeV.
Enforcing the No-Scale boundary conditions, including $B_\mu(M_{\cal F})=0$ on the Higgs bilinear soft
term, with the precision 7-year WMAP value on the dark matter relic density isolates
a highly constrained ``Golden Point'' located 
near $M_{1/2} = 455$~GeV and $\tan \beta = 15$ in the $\tan\beta-M_{1/2}$ plane,
which simultaneously satisfies all known experiments, and moreover corresponds to
an imminently observable proton decay rate.
Because the universal gaugino mass is actually determined 
from established low energy data via Renormalization Group Equation (RGE) running, 
there are no surviving arbitrary scale parameters in the present model.

\end{abstract}


\preprint{ACT-10-10, MIFPA-10-35}

\maketitle


{\bf Introduction~--}
The driving aim of theoretical physics is to achieve maximal efficiency 
in the correlation of observations.
This entails the unification of apparently distinct forces under 
a master symmetry group, and the
successful reinterpretation of experimentally constrained 
parameters and finely tuned scales
as dynamically evolved consequences of the underlying equations of motion.

We propose in this paper a variation of the 
No-Scale Supergravity~\cite{Cremmer:1983bf} scenario which successfully eliminates
all extraneously adjustable degrees of freedom while dynamically 
addressing all fundamental scales
and maintaining consistency with all low energy phenomenology constraints, 
including the precision electroweak scale data~\cite{:2009ec},
the 7-year WMAP constraint on dark matter relic density~\cite{Komatsu:2010fb},
the experimental limits on 
the Flavor Changing Neutral Current (FCNC) 
process $b \rightarrow s\gamma$~\cite{Barberio:2007cr, Misiak:2006zs},
the anomalous magnetic moment of the muon~\cite{Bennett:2004pv}, 
the process $B_{s}^{0} \rightarrow \mu^+ \mu^-$~\cite{:2007kv}, and
the LEP limit on the lightest CP-even Higgs boson 
mass~\cite{Barate:2003sz}, additionally predicting 
an experimentally safe,
yet still imminently observable proton lifetime.

The No-Scale picture inherits an associative weight of motivation from
 its robustly generic
and natural appearance across string-, M-, and ${\cal F}$-theory 
derived model building efforts~\cite{Witten:1985xb, Li:1997sk}.
It represents moreover a case study in reductionism, wherein the
 universal scalar mass $M_0$,  universal trilinear soft term $A$
and Higgs bilinear soft term $ B_{\mu}$ each vanish at
 some common high mass boundary,
and only the single universal gaugino mass parameter $M_{1/2}$ is left to float free.
All low energy scales are dynamically generated by quantum corrections,
{\it i.e.} running under the RGEs, to the classically flat potential.

This appealing perspective however, has historically been 
undermined by a basic inconsistency of
the $M_0 = 0$ condition as applied at a GUT scale of 
order $10^{16}$~GeV with precision phenomenology.
Attempts~\cite{Ellis:2001kg, Schmaltz:2000gy,Ellis:2010jb}
to reinterpret the No-Scale paradigm as a boundary near 
the Planck scale have met with
some exciting success, but we suggest that these efforts 
have been missing one most crucial piece
of the puzzle.  Our prior study, 
succinctly dubbed ${\cal F}$-$SU(5)$~\cite{Jiang:2006hf, Jiang:2009zza, Li:2010dp},
of the ${\cal F}$-lipped $SU(5)$ GUT~\cite{F-SU5, Lopez:1992kg}
supplemented by ${\cal F}$-theory derived vector-like multiplets 
at the TeV scale, provides the essential
rationale for the separation of an initial unification 
of the $SU(3)_C \times SU(2)_L $ gauge symmetry
at a mass $M_{32}$ near the traditional GUT scale, from a second phase
running up to a point $M_{\cal F}$ of final unification near the reduced Planck
mass~\cite{Jiang:2006hf, Jiang:2009zza}.
The dual high scales of ${\cal F}$-$SU(5)$ fit hand to glove
with the proposal for salvaging the no-scale conditions.

Only a small portion of viable
parameter space appears to be consistent with 
the $B_{\mu}(M_{\cal F}) = 0$ condition, which thus
constitutes a strong constraint. In the narrow 
region of overlap, we identify a highly
confined ``Golden Point'' at which all phenomenological limits are respected.
Moreover, since the boundary value of the universal gaugino mass $M_{1/2}$, and
even the unification scale $M_{\cal F}$ itself, are
established by the low energy experiments via
RGE running, we are not left with
any surviving scale parameters in the present model.

{\bf No Scale Supergravity~--}
Supersymmetry (SUSY) naturally solves
the gauge hierarchy problem in the Standard
Model (SM), and suggests, along with $R$ parity conservation,
the lightest supersymmetric particle (LSP)
as a suitable cold dark matter candidate.
Since we do not see mass degeneracy of the superpartners however,
SUSY must be broken around the TeV scale. In GUTs with
gravity mediated supersymmetry breaking, called
the supergravity models, 
we can fully characterize the supersymmetry breaking 
soft terms by four universal parameters
(gaugino mass $M_{1/2}$, scalar mass $M_0$, trilinear soft term $A$, and
the low energy ratio of Higgs vacuum expectation values (VEVs) $\tan\beta$),
plus the sign of the Higgs bilinear mass term $\mu$.

No-Scale Supergravity was proposed~\cite{Cremmer:1983bf},
to address the cosmological flatness problem,
as the subset of supergravity models
which satisfy the following three constraints:
(i) The vacuum energy vanishes automatically due to the suitable
 K\"ahler potential; (ii) At the minimum of the scalar
potential, there are flat directions which leave the 
gravitino mass $M_{3/2}$ undetermined; (iii) The quantity
${\rm Str} {\cal M}^2$ is zero at the minimum. If the third condition
were not true, large one-loop corrections would force $M_{3/2}$ to be
either identically zero or of the Planck scale. A simple K\"ahler potential which
satisfies the first two conditions is~\cite{Cremmer:1983bf}
\bea 
K &=& -3 {\rm ln}( T+\overline{T}-\sum_i \overline{\Phi}_i
\Phi_i)~,~
\label{NS-Kahler}
\eea
where $T$ is a modulus field and $\Phi_i$ are matter fields.
The third condition is model dependent and can always be satisfied in
principle~\cite{Ferrara:1994kg}.
For the simple K\"ahler potential in Eq.~(\ref{NS-Kahler})
we automatically obtain the no-scale boundary condition
$M_0=A=B_{\mu}=0$ while $M_{1/2}$ is allowed,
and indeed required for SUSY breaking.
Because the minimum of the electroweak (EW) Higgs potential 
$(V_{EW})_{min}$ depends on $M_{3/2}$,  the gravitino mass is 
determined by the equation $d(V_{EW})_{min}/dM_{3/2}=0$.
Thus, the supersymmetry breaking scale is determined 
dynamically. No-scale supergravity can be
realized in the compactification of the weakly coupled
heterotic string theory~\cite{Witten:1985xb} and the compactification of
M-theory on $S^1/Z_2$ at the leading order~\cite{Li:1997sk}. 

{\bf Models~--}
In the flipped $SU(5)$ GUT~\cite{F-SU5}
there are three families of SM fermions 
whose quantum numbers under the $SU(5)\times U(1)_{X}$ gauge group are
\bea
F_i={\mathbf{(10, 1)}},~ {\bar f}_i={\mathbf{(\bar 5, -3)}},~
{\bar l}_i={\mathbf{(1, 5)}},
\label{smfermions}
\eea
where $i=1, 2, 3$. 

To break the GUT and electroweak gauge symmetries, we 
introduce two pairs of Higgs fields
\bea
H={\mathbf{(10, 1)}},~{\overline{H}}={\mathbf{({\overline{10}}, -1)}},
~h={\mathbf{(5, -2)}},~{\overline h}={\mathbf{({\bar {5}}, 2)}}.
\label{Higgse1}
\eea

To separate the $M_{32}$ and $M_{\cal F}$ scales
and obtain true string-scale gauge coupling unification in 
free fermionic string models~\cite{Jiang:2006hf, Lopez:1992kg} or
the decoupling scenario in F-theory models~\cite{Jiang:2009zza},
we introduce vector-like particles which form complete
flipped $SU(5)\times U(1)_X$ multiplets.
In order to avoid the Landau pole
problem for the strong coupling constant, we can only introduce the
following two sets of vector-like particles around the TeV 
scale~\cite{Jiang:2006hf}
\begin{eqnarray}
&& Z1:  XF ={\mathbf{(10, 1)}}~,~
{\overline{XF}}={\mathbf{({\overline{10}}, -1)}}~;~\\
&& Z2: XF~,~{\overline{XF}}~,~Xl={\mathbf{(1, -5)}}~,~
{\overline{Xl}}={\mathbf{(1, 5)}}
~.~\,
\end{eqnarray}
In this paper, we only consider the flipped
$SU(5)\times U(1)_X$ models with 
$Z2$ set of vector-like particles.
The discussions for the models with 
$Z1$ set and heavy threshold corrections~\cite{Jiang:2009zza}
are similar.



{\bf The Golden Point~--}
In the No-Scale context, we impose $M_0 = A = B_\mu$ = 0 at the unification scale
$M_{\cal F}$, and allow distinct inputs for the single parameter $M_{1/2}(M_{\cal F})$ to 
translate under the RGEs to distinct low scale outputs of $B_\mu$ and the Higgs mass-squares $M^2_{H_u}$
and $M^2_{H_d}$.  This continues until the point of spontaneous breakdown of the electroweak symmetry at
$M^2_{H_u} + \mu^2 = 0$, at which point minimization of the broken potential establishes the
physical low energy values of $\mu$ and $\tan \beta$.
In practice however, this procedure is at odds with the existing
{\tt SuSpect 2.34} code~\cite{Djouadi:2002ze} base from which our primary routines have been adapted.
In order to impose the minimal possible refactoring, we have instead
opted for an inversion wherein $M_{1/2}$ and $\tan \beta$ float
as two effective degrees of freedom. Thus, we do not fix $B_\mu(M_{\cal F})$.
We take $\mu >0$ as suggested by the results of $g_{\mu}-2$ for the muon,
and use 1 TeV for the universal vector-like particle mass~\cite{Li:2010mi}.

The relic LSP neutralino density, WIMP-nucleon direct 
detection cross sections  and 
photon-photon annihilation cross sections
are computed with {\tt MicrOMEGAs 2.1}~\cite{Belanger:2008sj} 
wherein the revised {\tt SuSpect} RGEs have also implemented.
We use a top quark mass of $m_{t}$ = 173.1 GeV~\cite{:2009ec} and 
employ the following experimental constraints:
(1) The WMAP 7-year measurements of 
the cold dark matter density~\cite{Komatsu:2010fb}, 
0.1088 $\leq \Omega_{\chi} \leq$ 0.1158. We allow $\Omega_{\chi}$ to be
larger than the upper bound due to a possible $\cal{O}$(10) 
dilution factor~\cite{Mavromatos:2009pm}
and to be smaller than the lower bound due to multicomponent
dark matter. (2) The experimental limits on 
the FCNC process, $b \rightarrow s\gamma$. 
We use the limits 
$2.86 \times 10^{-4} \leq Br(b \rightarrow s\gamma) 
\leq 4.18 \times 10^{-4}$~\cite{Barberio:2007cr, Misiak:2006zs}.
(3) The anomalous magnetic moment of the muon, $g_{\mu} - 2$. 
We use the $2\sigma$ level boundaries, 
$11 \times 10^{-10} < \Delta a_{\mu} < 44 \times 10^{-10}$~\cite{Bennett:2004pv}. 
(4) The process $B_{s}^{0} \rightarrow \mu^+ \mu^-$ where we take the upper bound to be
 $Br(B_{s}^{0} \rightarrow \mu^{+}\mu^{-}) < 5.8 \times 10^{-8}$~\cite{:2007kv}. 
(5) The LEP limit on the lightest CP-even Higgs boson 
mass, $m_{h} \geq 114$ GeV~\cite{Barate:2003sz}.

\begin{figure}[ht]
	\centering
		\includegraphics[width=0.46\textwidth]{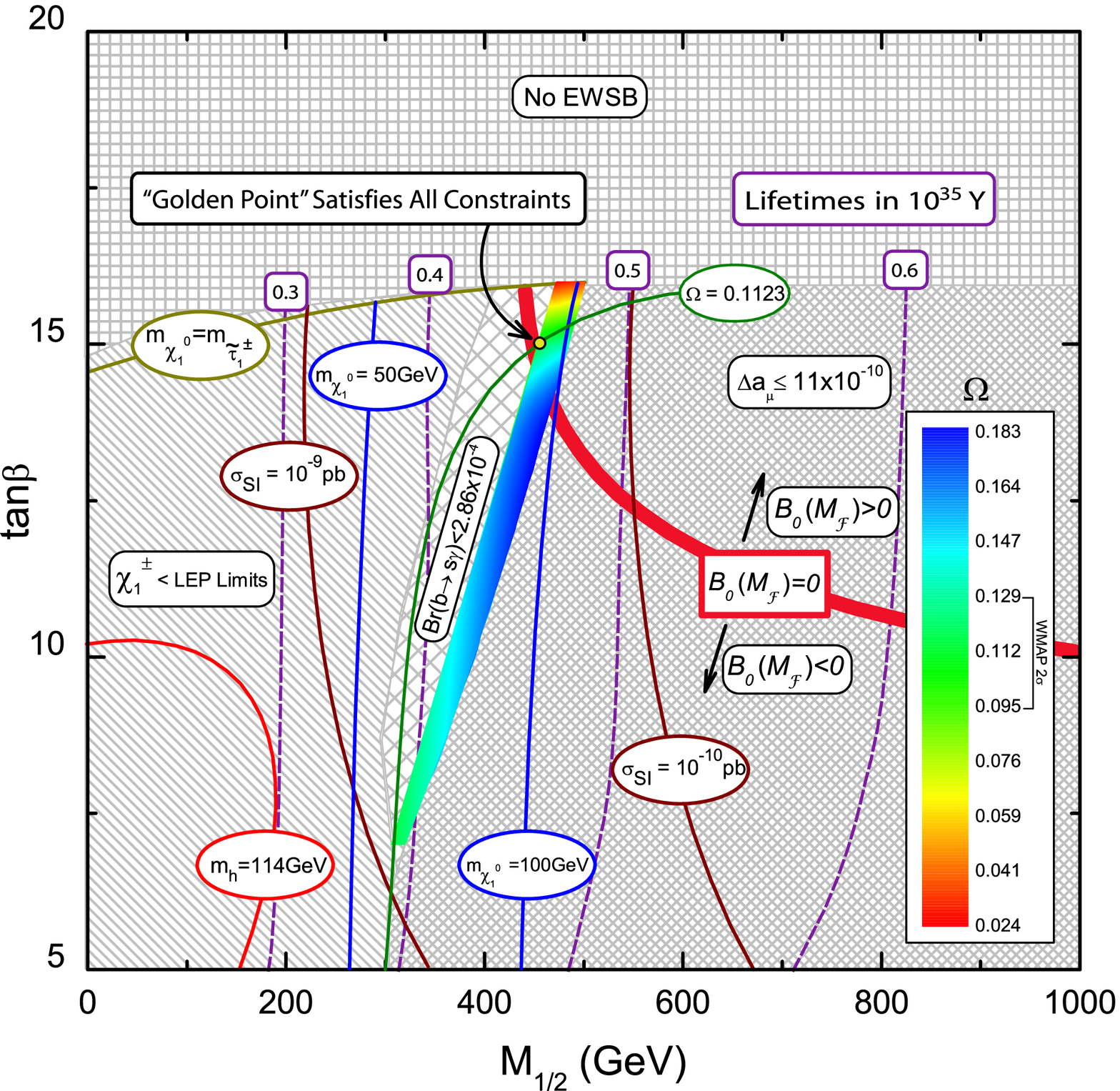}
		\caption{Viable parameter space in the $\tan\beta-M_{1/2}$ plane.
The Golden Point is annotated. 
The thin, dark green line denotes the WMAP 7-year central value $\Omega_{\chi} = 0.1123$. The dashed purple contours
label $p \!\rightarrow\! {(e\vert\mu)}^{\!+}\! \pi^0$ proton lifetime 
predictions, in units of $10^{35}$ years.}
	\label{fig:NoScale_M12_vs_tanb}
\end{figure}

In the $\tan\beta-M_{1/2}$ plane,
$B_\mu(M_{\cal F})$ is then calculated along
with the low energy supersymmetric particle
 spectrum and checks on various experimental constraints.
The subspace corresponding to a No-Scale model is clearly
then a one dimensional slice of this manifold, as demonstrated in Fig.~\ref{fig:NoScale_M12_vs_tanb}.
It is quite remarkable that
the $B_\mu(M_{\cal F}) = 0$ contour so established runs sufficiently perpendicular to the WMAP
strip that the point of intersection effectively absorbs our final degree
of freedom, creating what we have labeled as a No-Parameter Model.  It is truly
extraordinary however that this intersection occurs exactly at the centrally
preferred relic density, that being our strongest experimental constraint.  We emphasize
again that there did not have to be an experimentally viable $B_\mu(M_{\cal F}) = 0$
solution, and that the consistent realization of this scenario depended crucially
on several uniquely identifying characteristics of the underlying proposal.  Specifically
again, it appears that the No-Scale condition comes into its own only when applied at
near the Planck mass, and that this is naturally identified as
the point of the final ${\cal F}$-$SU(5)$ unification, which is naturally extended and
decoupled from the primary GUT scale only via the modification to the RGEs from the
TeV scale ${\cal F}$-theory vector-like multiplet content.  The union of our
top-down model based constraints with the bottom-up experimental data exhausts
the available freedom of parameterization in a uniquely consistent and predictive manner,
phenomenologically defining a truly Golden Point
near $M_{1/2} = 455$ GeV and $\tan\beta$ = 15.


\begin{table}[ht]
  \small
	\centering
	\caption{Spectrum (in GeV) for the Golden Point in Fig.~\ref{fig:NoScale_M12_vs_tanb}. 
Here, $\Omega_{\chi}$ = 0.1123, $\sigma_{SI} = 1.9 \times 10^{-10}$ pb, and
$\left\langle \sigma v \right\rangle_{\gamma\gamma} = 1.7 \times 10^{-28} ~cm^{3}/s$.
The central prediction for the $p \!\rightarrow\! {(e\vert\mu)}^{\!+}\! \pi^0$ 
proton lifetime is $4.6 \times 10^{34}$ years.}
		\begin{tabular}{|c|c||c|c||c|c||c|c||c|c||c|c|} \hline		
    $\widetilde{\chi}_{1}^{0}$&$95$&$\widetilde{\chi}_{1}^{\pm}$&$185$&$\widetilde{e}_{R}$&$150$&$\widetilde{t}_{1}$&$489$&$\widetilde{u}_{R}$&$951$&$m_{h}$&$120.1$\\ \hline
    $\widetilde{\chi}_{2}^{0}$&$185$&$\widetilde{\chi}_{2}^{\pm}$&$826$&$\widetilde{e}_{L}$&$507$&$\widetilde{t}_{2}$&$909$&$\widetilde{u}_{L}$&$1036$&$m_{A,H}$&$920$\\ \hline
    
    $\widetilde{\chi}_{3}^{0}$&$821$&$\widetilde{\nu}_{e/\mu}$&$501$&$\widetilde{\tau}_{1}$&$104$&$\widetilde{b}_{1}$&$859$&$\widetilde{d}_{R}$&$992$&$m_{H^{\pm}}$&$925$\\ \hline
    $\widetilde{\chi}_{4}^{0}$&$824$&$\widetilde{\nu}_{\tau}$&$493$&$\widetilde{\tau}_{2}$&$501$&$\widetilde{b}_{2}$&$967$&$\widetilde{d}_{L}$&$1039$&$\widetilde{g}$&$620$\\ \hline
		\end{tabular}
		\label{tab:masses}
\end{table}

\begin{figure}[ht]
	\centering
		\includegraphics[width=0.46\textwidth]{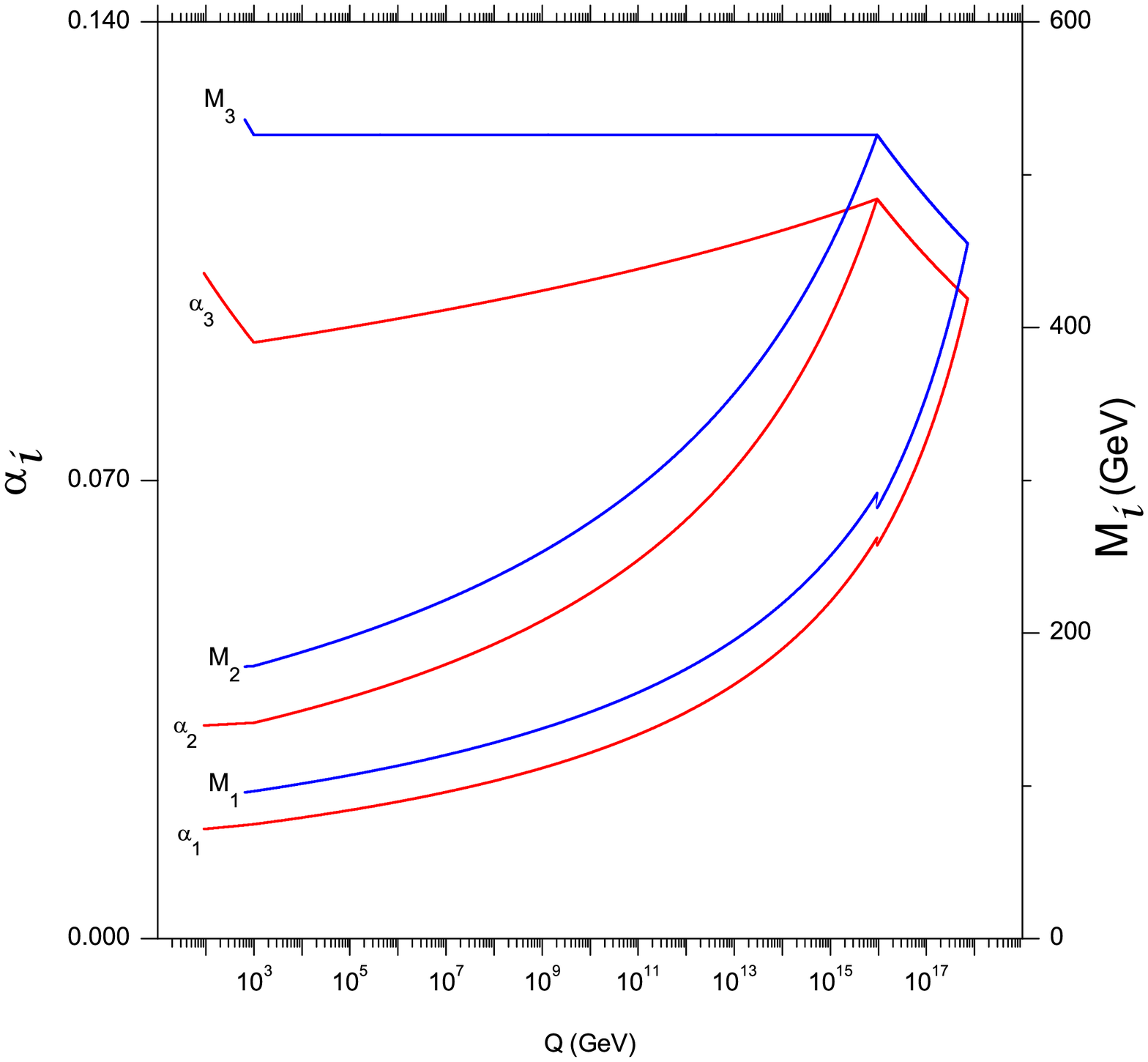}
		\caption{ RGE Running of the SM gauge couplings and gaugino masses from 
the EW scale to the unification scale $M_{\cal F}$. Notice the discontinuity of $U(1)_{Y}$ as it remixes between the $U(1)_{X}$ and that which emerges out of broken SU(5) at the scale $M_{32} \simeq 1 \times 10^{16}$ GeV. Advancing from this interim stage, $SU(5) \times U(1)_{X}$ is unified at a higher scale $M_{\cal F} \simeq 7 \times 10^{17}$ GeV.}
	\label{fig:NoScale_alpha}
\end{figure}

The Golden Point parameters are 
$M_{1/2} = 455.3$ GeV, $\tan \beta = 15.02$, and
the point is in full compliance with the WMAP 7-year 
results with $\Omega_{\chi}$ = 0.1123. 
It also satisfies the CDMSII~\cite{Ahmed:2008eu},
Xenon100~\cite{Aprile:2010um}, and
FERMI-LAT space telescope constraints~\cite{Abdo:2010dk}, with $\sigma_{SI} = 1.9 \times 10^{-10}$ pb and 
$\left\langle \sigma v \right\rangle_{\gamma\gamma} = 1.7 \times 10^{-28} ~cm^{3}/s$. 
The proton lifetime is about $4.6\times 10^{34}$ years, which
is well within reach of the upcoming Hyper-Kamiokande~\cite{Nakamura:2003hk}
and DUSEL~\cite{DUSEL} experiments. Inspecting the supersymmetric particle and Higgs spectrum
for the Golden Point in Table~\ref{tab:masses} reveals that the additional 
contribution of the 1 TeV vector-like particles lowers the gluino mass quite 
dramatically. The gluino mass $M_{3}$ runs flat from the $M_{32}$ unification 
scale to 1 TeV as shown in Fig.~\ref{fig:NoScale_alpha}, though, 
due to supersymmetric radiative corrections, the physical gluino mass at the EW scale 
is larger than $M_{3}$ at the $M_{32}$ scale. This is true 
for the full parameter space. For the Golden Point, 
the LSP neutralino is 99.8\% Bino. Similarly to the mSUGRA picture,
the point is in
the stau-neutralino coannihilation region, but the gluino is lighter
than the squarks in our models, with the exception of the lightest stop.

\begin{figure}[ht]
		\includegraphics[width=0.46\textwidth]{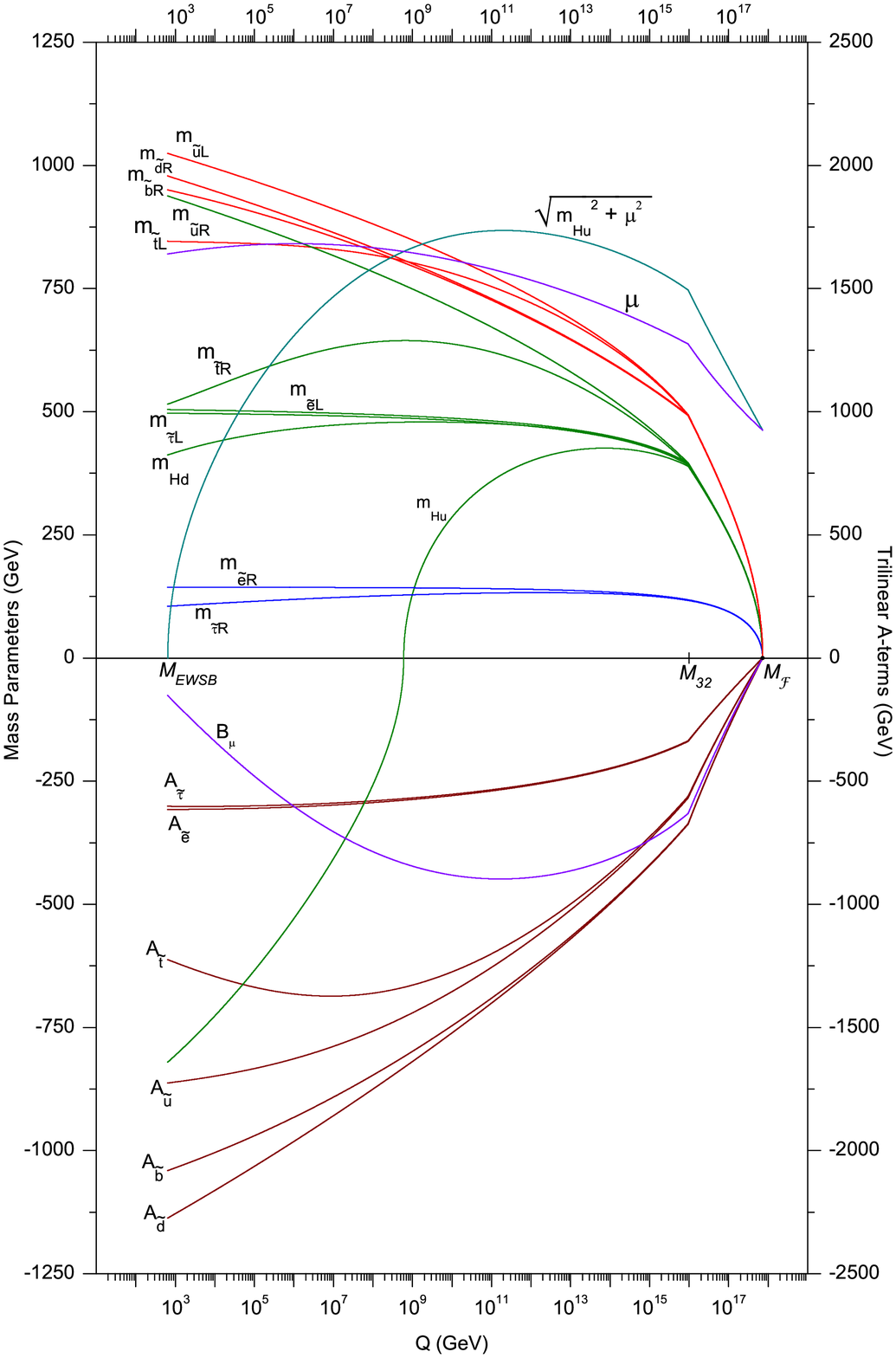}
		\caption{RGE Running of the $\mu$ term and SUSY breaking
soft terms from the EW scale to the unification scale $M_{\cal F}$.}
	\label{fig:NoScale_parameters}
\end{figure}

We plot gauge coupling and gaugino mass unification for the Golden 
Point in Fig.~\ref{fig:NoScale_alpha}.
The figure explicitly demonstrates the two-step unification of flipped $SU(5)\times U(1)_X$. In this work, we consider the two-loop RGE running for the gauge couplings, however, we only consider the one-loop RGE running for the gaugino masses. In ${\cal F}$-$SU(5)$ models, the one-loop beta function for $SU(3)_{C}$ is zero due to the vector-like particle contributions. Therefore, as shown in Fig.~\ref{fig:NoScale_alpha}, $M_{3}$ is constant from the electroweak scale to the $M_{32}$ scale since the beta coefficient $b_{3}$ = 0. In contrast, the gauge couplings and gaugino masses for the $SU(2)_{L} \times U(1)_{Y}$ gauge symmetry track each other in Fig.~\ref{fig:NoScale_alpha} since the gauge couplings for $SU(2)_{L} \times U(1)_{Y}$ are weak, thus the two-loop effects are small.
In addition, we present the RGE running for the $\mu$ term,
the SUSY breaking scalar masses, trilinear A-terms, and bilinear $B_{\mu}$ term
in Fig.~\ref{fig:NoScale_parameters}.
Note in particular that the EW symmetry breaking occurs when $H_{u}^{2} + \mu^{2}$ goes negative.

{\bf Conclusion~--}~We have studied No-Scale Supergravity
in the context of a ${\cal F}$-lipped $SU(5)\times U(1)_X$ GUT 
supplemented with ${\cal F}$-theory derived TeV-scale vector-like particles.
With the no-scale boundary condition applied at the point of final unification
$M_{\cal F}$, we find a very small ``Golden Point'' of viable parameter space
that is consistent with all known experiments, while fixing
all extraneous model parameters.
For the Golden Point, we have discussed
unification of the SM gauge couplings and 
gaugino masses, and the RGE running for 
the $\mu$ term and supersymmetry breaking
soft terms.
Proton decay predictions are well within the range accessible
to the future Hyper-Kamiokande and DUSEL experiments.
Because the universal gaugino mass is determined 
by the low energy known experiments via RGE running,
we emphasize that there are no 
surviving arbitrary scale parameters in this model.

{\bf Acknowledgments~--}~This research was supported in part 
by  the DOE grant DE-FG03-95-Er-40917 (TL and DVN),
by the Natural Science Foundation of China 
under grant No. 10821504 (TL),
and by the Mitchell-Heep Chair in High Energy Physics (TL).


\end{document}